\documentclass[preprint2,times]{aastex631}

\PassOptionsToPackage{dvipsnames,svgnames,x11names}{xcolor}
\usepackage{tabularx}

\usepackage{amsmath}  
\usepackage{amssymb}  
\usepackage[T1]{fontenc}
\usepackage{xspace}
\usepackage{multirow}
\usepackage{xcolor}

\usepackage{booktabs}

\defcitealias{kenworthy_salt3_2021}{K21}

\usepackage{newunicodechar,graphicx}
\DeclareRobustCommand{\okina}{%
  \raisebox{\dimexpr\fontcharht\font`A-\height}{%
    \scalebox{0.8}{`}%
  }%
}
\newunicodechar{ʻ}{\okina}

\newif{\ifchangetext}
\changetextfalse

\InputIfFileExists{\jobname-options}

\ifchangetext
  
  \newcommand{\changenote}[1]{\textcolor{blue}{ \bf #1}}x
\else
  
  \newcommand{\changenote}[1]{}
\fi

\hypersetup{
    colorlinks=True,
    citecolor=black,
    linkcolor=black,
    filecolor=magenta,      
    urlcolor=cyan,
}

\def\arcsec{\ensuremath{^{\prime\prime}}}

\def\xone{\ensuremath{x_{1}}}
\def\C{\ensuremath{\mathcal{C}}}

\newcommand{\hst}{\textit{HST}\xspace}

\newcommand{\webb}{\textit{JWST}\xspace}

\newcommand{\STScI}{Space Telescope Science Institute, Baltimore, MD 21218, USA}
\newcommand{\Steward}{Steward Observatory, University of Arizona, 933 N. Cherry Avenue, Tucson, AZ 85721, USA}
\newcommand{\JHU}{Physics and Astronomy Department, Johns Hopkins University, Baltimore, MD 21218, USA}

\newcommand{\NEF}{NASA Einstein Fellow}
\newcommand{\ISEF}{ISEF International Fellowship}
\newcommand{\Cavendish}{Cavendish Laboratory, University of Cambridge, 19 JJ Thomson Avenue, Cambridge, CB3 0HE, UK}
\newcommand{\KavliCambridge}{Kavli Institute for Cosmology, University of Cambridge, Madingley Road, Cambridge CB3 0HA, UK}

\newcommand{\snz}{$3.613 \pm 0.001 $\xspace}
\newcommand{\jadesarea}{$25$}
\newcommand{\jadesdepth}{$30$}
\newcommand{\mic}{\ensuremath{\mu\rm m}\xspace}

\newcommand{\hostmass}{$8.41^{+0.12}_{-0.12}$}
\newcommand{\hostage}{$8.55^{+0.15}_{-0.17}$}

\newcommand{\hostextinction}{$0.15^{+0.11}_{-0.07}$}
\newcommand{\hostgasmetallicitylogOH}{$8.1\pm 0.2$}
\newcommand{\hostgasmetallicitysolar}{$0.3\pm 0.1$}
\newcommand{\hostSFR}{$0.31^{+0.08}_{-0.06}$}

\newcommand{\highzIIcoord}{R.A.=$3$h$32$m$39.4574$s decl.=$-27$d$50$m$19.6660$s\xspace}

\usepackage{xspace}

\newcommand{\highzII}{AT~2023adsv\xspace}
\newcommand{\highzIIhost}{JADES-GS+53.16439-27.83877\xspace}
\usepackage[mathlines]{lineno}
\begin{document}

\title{Discovery of a likely Type II SN at $z$=3.6 with \webb}

\author[0000-0003-4263-2228
]{D.~A.~Coulter}
\correspondingauthor{D.~A.~Coulter} 
\email{dcoulter@stsci.edu}
\affiliation{\STScI}

\author[0000-0002-2361-7201]{J.~D.\ R.\ Pierel}
\altaffiliation{\NEF}
\affiliation{\STScI}

\author[0000-0002-4781-9078]{C.~DeCoursey}
\affiliation{\Steward}

\author[0000-0003-1169-1954]{T.~J.~Moriya}
\affiliation{National Astronomical Observatory of Japan, National Institutes of Natural Sciences, 2-21-1 Osawa, Mitaka, Tokyo 181-8588, Japan}
\affiliation{Graduate Institute for Advanced Studies, SOKENDAI, 2-21-1 Osawa, Mitaka, Tokyo 181-8588, Japan}
\affiliation{School of Physics and Astronomy, Monash University, Clayton, VIC 3800, Australia}

\author[0000-0003-2445-3891]{M.~R.~Siebert}
\affiliation{\STScI}

\author[0000-0002-7593-8584]{B.~A.~Joshi}
\affiliation{\JHU}

\author[0000-0003-0209-674X]{M.~Engesser}
\affiliation{\STScI}

\author[0000-0002-4410-5387]{A.~Rest} 
\affiliation{\STScI}
\affiliation{\JHU}

\author[0000-0003-1344-9475]{E.~Egami}
\affiliation{\Steward}

\author[0000-0002-9301-5302]{M.~Shahbandeh} 
\affiliation{\STScI}

\author[0000-0003-1060-0723]{W.~Chen} 
\affiliation{Department of Physics,
Oklahoma State University, 145 Physical Sciences Bldg, Stillwater, OK
74078, USA}

\author[0000-0003-2238-1572]{O.~D.~Fox} 
\affiliation{\STScI}

\author[0000-0002-7756-4440]{L.~G.~Strolger} 
\affiliation{\STScI}

\author[0000-0002-0632-8897]{Y.~Zenati}
\altaffiliation{\ISEF}
\affiliation{\JHU}
\affiliation{\STScI}

\author[0000-0002-8651-9879] {A.~J.~Bunker}
\affiliation{Department of Physics, University of Oxford, Denys Wilkinson Building, Keble Road, Oxford OX1 3RH, UK}

\author[0000-0002-1617-8917] {P.~A.~Cargile}
\affiliation{Center for Astrophysics $|$ Harvard \& Smithsonian, 60 Garden St., Cambridge MA 02138 USA}

\author[0000-0002-2678-2560] {M.~Curti}
\affiliation{European Southern Observatory, Karl-Schwarzschild-Strasse 2, 85748 Garching, Germany}

\author[0000-0002-2929-3121] {D.~J.~Eisenstein}
\affiliation{Center for Astrophysics $|$ Harvard \& Smithsonian, 60 Garden St., Cambridge MA 02138 USA}

\author[0000-0003-3703-5154]{S.~Gezari}
\affiliation{\STScI}
\affiliation{\JHU}

\author[0000-0001-6395-6702]{S.~Gomez}
\affiliation{Center for Astrophysics $|$ Harvard \& Smithsonian, 60 Garden St., Cambridge MA 02138 USA}

\author[0000-0002-5063-0751]{M.~Guolo}
\affiliation{\JHU}

\author[0000-0003-4565-8239]{K.~Hainline}
\affiliation{\Steward}

\author[0000-0001-5754-4007] {J.~Jencson}
\affiliation{IPAC, Mail Code 100-22, Caltech, 1200 E.\ California Boulevard, Pasadena, CA 91125, USA}

\author[0000-0002-9280-7594] {B.~D.~Johnson}
\affiliation{Center for Astrophysics $|$ Harvard \& Smithsonian, 60 Garden St., Cambridge MA 02138 USA}

\author[0000-0003-2495-8670]{M.~Karmen}
\affiliation{\JHU}

\author[0000-0002-4985-3819]{R.~Maiolino}
\affiliation{\KavliCambridge}
\affiliation{\Cavendish}
\affiliation{Department of Physics and Astronomy, University College London, Gower Street, London WC1E 6BT, UK}

\author[0000-0001-9171-5236]{R.~M.~Quimby}
\affiliation{Department of Astronomy/Mount Laguna Observatory, San Diego State University, 5500 Campanile Drive, San Diego, CA 92812-1221, USA}
\affiliation{Kavli Institute for the Physics and Mathematics of the Universe (WPI), The University of Tokyo Institutes for Advanced Study, The University of Tokyo, Kashiwa, Chiba 277-8583, Japan}

\author[0000-0002-5104-8245]{P.~Rinaldi}
\affiliation{\Steward}

\author[0000-0002-4271-0364] {B.~Robertson}
\affiliation{Department of Astronomy and Astrophysics, University of California, Santa Cruz, 1156 High Street, Santa Cruz CA 96054, USA}

\author[0000-0002-8224-4505]{S.~Tacchella}
\affiliation{\KavliCambridge}
\affiliation{\Cavendish}

\author[0000-0002-4622-6617]{F.~Sun}
\affiliation{Center for Astrophysics $|$ Harvard \& Smithsonian, 60 Garden St., Cambridge MA 02138 USA}

\author[0000-0001-5233-6989]{Q.~Wang} 
\affiliation{Department of Physics and Kavli Institute for Astrophysics and Space Research, Massachusetts Institute of Technology, 77
Massachusetts Avenue, Cambridge, MA 02139, USA}

\author[0000-0002-4043-9400]{T.~Wevers}
\affiliation{\STScI}

\begin{abstract}
Transient astronomy in the early, high-redshift ($z>3$) Universe is an unexplored regime that offers the possibility of probing the first stars and the Epoch of Reionization. During Cycles 1 and 2 of the \textit{James Webb Space Telescope} (\textit{JWST}), the \webb Advanced Deep Extragalactic Survey (JADES) program enabled one of the first searches for transients in deep images ($\sim$\jadesdepth~AB~mag) over a relatively wide area (\jadesarea~arcmin$^2$). One transient, \highzII, was discovered with an F200W magnitude of $28.04$~AB~mag, and subsequent \webb observations revealed that the transient is a likely supernova (SN) in a host with $z_{\rm spec}=$~\snz
and an inferred metallicity at the position of the SN of Z$_{*}$~=~\hostgasmetallicitysolar~Z$_\odot$. At this redshift, the first detections in F115W and F150W show that \highzII had bright rest-frame ultraviolet flux at the time of discovery. The multi-band light curve of \highzII is best matched by a template of an SN~IIP with a peak absolute magnitude of $M_{B}\approx-18.3$~AB mag.
We find a good match to a $20M_\odot$ red supergiant progenitor star with an explosion energy of $2.0\times 10^{51}$ ergs, likely higher than normally observed in the local Universe, but consistent with SNe~IIP drawn from local, lower metallicity environments. \highzII is the most distant photometrically classified SN~IIP yet discovered with a spectroscopic redshift measurement, and may represent a global shift in SNe~IIP properties as a function of redshift. 
\end{abstract}
\keywords{supernovae: individual (\highzII); SN~II - infrared: supernovae - stars: massive - galaxies: abundances}

\section{Introduction}
\label{sec:intro}

Core-collapse supernovae (CCSNe) are the explosive deaths of massive stars with initial masses $>8$~M$_{\odot}$ and are remarkably diverse in their properties \citep{OppenheimerSnyder39, Kobulnicky_Skillman97, Vanbeveren+98A, Heger2003, Smartt09, Dessart2020, BurrowsVartanyan21Nat}. This diversity is driven by the broad range of their progenitor masses, which sensitively affect their evolution, setting the initial conditions for both their stellar structure and circumstellar environments prior to collapse and resulting in a similarly broad range of explosion energies, ejecta compositions and observed luminosities \citep{SmithN14, Gal-yam+14, WuFuller21}. These explosions connect to astrophysical phenomena across many scales --- due to their high mass, the progenitors of CC\,SNe have short lifetimes and therefore trace the instantaneous star formation rate (SFR) of their locales; their rates constrain the high-mass end of the Initial Mass Function (IMF); their explosions deposit energy and momentum into the interstellar medium (ISM) providing a feedback mechanism to moderate star formation; they enrich the ISM with metals and are factories for cosmic dust; and they produce ionizing photons that contribute to the reionization of the Universe.

CC\,SNe, and in particular SNe\,II, are in principle luminous enough to be observed at cosmological distances, making them intriguing probes of the early Universe. However, their peak emission in optical bands is shifted into the infrared (IR) at high-redshift. In the last two decades, work based on the {\em Hubble Space Telescope} (\hst) has pushed the study of CCSNe rates and properties to further distances \citep{Botticella2008, Bazin2009, Graur2011, Melinder2012, Dahlen2012}, culminating with observations from the Cosmic Assembly Near-infrared Deep Extragalactic Legacy Survey \citep[CANDELS;][]{Grogin2011, Koekemoer2011} and Cluster Lensing And Supernova survey with Hubble \citep[CLASH;][]{Postman2012}, which constrained the CCSN rate out to $z \approx 2.5$ \citep{Strolger2015}. 

CC\,SNe discovered at even greater distances ($z>2.5$) will peak in at wavelengths of $2$~\mic and beyond, placing more distant samples out of reach for \hst but not of the {\em James Webb Space Telescope} (\webb). Indeed, \webb is already removing this barrier to discovering distant and observer-frame IR bright CC\,SNe due to its combination of wavelength coverage and sensitivity, opening a new frontier in transient astronomy with the discovery of several high-$z$ SNe since its launch \citep{Chen2022, engesser_detection_2022,engesser_discovery_2022, decoursey_discovery_2023,
decoursey_discovery_2023-1,
decoursey_discovery_2023b,
decoursey_jades_2024,pierel_jwst_2024,pierel_lensed_2024,pierel_discovery_2024,pierel_testing_2024, Siebert2024}. Such discoveries are vital laboratories to test topics such as whether the CC\,SN rate follows the cosmic SFR density or if the high-mass end of the IMF flattens with redshift in low metallicity stellar populations \citep{Larson1998, Ziegler2022}. 

While metallicity could very plausibly impact the rate of CC\,SNe, it also impacts their massive stellar progenitors and, therefore, their explosive properties. In particular, the metallicity of the progenitors to SNe~II affects not only their mass loss \citep{Vink01, Mokiem2007}, but their internal structure and convective efficiency \citep{Heger2003, Dessart2013}, leading to a range of pre-explosion envelope masses, stellar radii, and the presence of circumstellar material (CSM). 
These, in turn, yield a diversity of observed SN~II properties, such as their resulting colors, peak luminosities, and for SNe~IIP, their plateau durations \citep{Sanyal2017, Dessart2013}. 

Indeed, recent studies of nearby SNe~II have found evidence of a link between the metallicity of the host galaxy and an SN II's spectral and photometric properties \citep{Anderson2016, Taddia2016, Gutierrez2017, Gutierrez2018}, and theoretical work has suggested that SNe~IIP may be effective probes of a host's underlying metallicity if properly calibrated \citep{Dessart2014}. Other empirical studies studies have shown that SNe~II tend to be up to $\sim0.5$ mag more luminous than the average SN~II at solar metallicities \citep{Tucker2024}, and that the depth of the `pseudo' equivalent width (pEW) of Fe II ($\lambda$5018~\AA) is suppressed at lower metallicity \citep{Scott2019}. These findings are potentially biased, however, by the fact that most of these SNe have been discovered within the local universe where hosts tend to be luminous and at solar metallicity. Local, notable exceptions include the discovery of SN~20015bs, which purports that its progenitor was particularly massive with a Zero Age Main-Sequence (ZAMS) mass of $\approx17-25M_\odot$ \citep{Anderson2018}, and SN~2023ufx, a luminous ($M_{g}\approx-18.5$~mag) SN~II discovered in a low-metallicity dwarf galaxy \citep{Tucker2024}; both SNe with metallicities of $Z\leq0.1Z_{\odot}$, and both potential analogues to SNe~II discovered at high-$z$.

In general, lower metallicities should lead to lower mass loss rates for SN\,II progenitors, yielding more massive progenitors \citep[barring interactions with binary companions;][]{Lamers1999, Kudritzki2000}, and the reduction in stellar envelope opacity should result in stars with smaller stellar radii \citep{Sanyal2017}. These effects may combine to produce progenitors with substantially higher rotation rates \citep{Woosley2006, Maeder2012} and potentially connect lower metallicity environments at high-$z$ with a diverse menagerie of exotic SN types including pair-instability SNe \citep[PISNe;][]{Kasen2011, Woosley2017}, superluminous supernovae \citep[SLSNe;][]{Quimby2011, GalYam2019}, and the supernovae associated with long gamma-ray bursts \citep[LGRB;][]{Zeh2004, Fruchter2006, Modjaz14}. However, these assertions need to be tested through the discovery of many more CC\,SNe in the early Universe.

Fortunately, this is a task to which \webb is particularly well-suited. The \textit{JWST} Advanced Deep Extragalactic Survey (JADES) program \citep{eisenstein_jades_2023} observed $\sim25$~arcmin$^2$ of the sky to depths of $m_{AB} \gtrapprox 30$ in $9$ NIRCam filters in two separate epochs, the first between September 29 and October 5, 2022, and the second between September 28 and October 3, 2023. 
These repeated observations allowed for these images to be 
subtracted to discover new transients beyond the redshift limitations of \hst, with a sensitivity for CCSNe to $z>4$. Using this $\sim1$~year baseline dozens of new transient objects were discovered \citep[][hereafter D24]{decoursey_jades_2024}, and here we present a candidate for one of the most distant SN\,II discovered to date: \highzII, a very blue and likely sub-solar metallicity SN IIP-like transient located at \highzIIcoord (although see \citet{Cooke2012} and \citet{Gomez2024} for previous high-$z$ SLSNe candidates).
\highzII is embedded in its host, \highzIIhost, with a spectroscopically confirmed redshift of $z_{\rm spec}=$\snz. 

In what follows, we describe the identification and analysis of \highzII, as well as a brief comparison to other SNe~IIP in the local universe.
This paper is structured as follows: in \S\ref{sec:obs}, we present a summary of the observations for this supernova, our reduction of the data, and obtaining \highzII's host redshift; in \S\ref{sec:analysis} we describe our classification of \highzII as a likely SN~II and present the properties of its host and model \highzII's light curve, in \S\ref{sec:discussion} we discuss \highzII in the context of a sample of local SNe~IIP, and in \S\ref{sec:conclusion} we conclude with a discussion on the prospects for building an SNe\,IIP sample at high redshift and its use as a metallicity probe of the Universe, as well as the implications of the new frontier enabled by \webb. Throughout this paper, we assume a standard $\Lambda$CDM cosmology with $H_0=70$km s$^{-1}$ Mpc$^{-1}$, $\Omega_m=0.315$.

\begin{figure*}[th!]
    \centering
    \includegraphics[width=\textwidth]{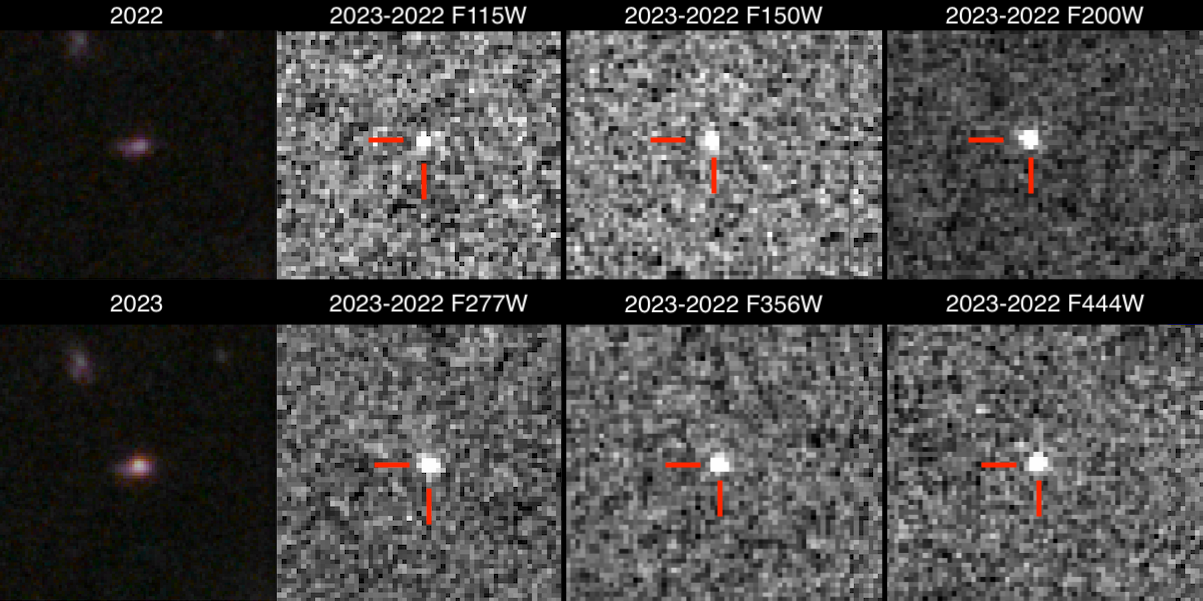}
    \caption{(Left column) Full color images using F115W+F150W (Blue) F200W+F277W (Green) and F356W+F444W (Red), with the $2022$ JADES epoch on top and $2023$ (including \highzII) on the bottom. (Column $2$-$4$) Difference images were created from the two JADES epochs ($2023-2022$), with the \highzII position marked with a red indicator. All images are drizzled to $0.03^{\prime\prime}/$pix and have the same spatial extent.}
    \label{fig:im_cutouts}
\end{figure*}

\section{Summary of Observations}
\label{sec:obs}
\highzII was discovered as a part of a transient search for the JADES program \citep{Eisenstein2023}, centered on the Great Observatories Origins Deep Survey's south field \citep[GOODS-S;][]{Giavalisco2004}. A full description of JADES, including its survey design, data products, the selection process for discovering new transients, and the follow-up observations of those subsequent discoveries through its approved DDT program, are described and presented in detail in D24.

To summarize, the first JADES observations were acquired between September $29$th, $2022$ and October $5$th, $2022$, in the NIRCam filters F090W, F115W, F150W, F200W, F277W, F335M, F356W, F410M, and F444W to a $5\sigma$ depth of m$_{AB}\sim30$. Nearly a year later, a second set of observations in the same filters and to the same depths were taken between September $29$th, $2023$ and October $3$rd, $2023$, resulting in an overlapping footprint of $\sim25$~arcmin$^2$ (both observations under PID 1180). During this second epoch, several observations failed, and subsets of the field were observed on November $15$th, $2023$, and January $1$st, $2024$. Upon the identification of many interesting transients in color, redshift, and luminosity space (see D24 for a complete accounting), a \webb Director's Discretionary Time (DDT) program was approved to follow up the most interesting transients in this field \citep{egami_jwst_2023}. These subsequent observations were obtained on November $28$th, $2023$ (NIRCam filters F115W, F150W, F200W, F277W, F356W, and F444W) and on January $1$st, $2024$ (NIRCam filters F150W, F200W, F277W, F356W, and F444W; PID 6541) with the latter epoch including NIRSpec multi-object spectroscopy (MOS) coverage using the micro-shutter assembly \citep[MSA;][]{Ferruit2022}. NIRSpec  
covered the most promising $\sim10$ transients (as well as a variety of galaxy spectra) 
using the MSA with the Prism (R$\sim100$) grating, of which \highzII was one, as well as two others that are described in a pair of companion papers \citep{Pierel2024c, Siebert2024}. Below, we describe the data reduction and photometric measurements we derive for \highzII.

\subsection{Measuring Photometry}
\label{sub:obs_phot}

\highzII is embedded in a relatively compact host and therefore obtaining accurate photometric measurements requires removing the underlying host light from the SN position. We achieve this through the process of ``difference imaging,'' or subtracting a template image (preferably with no SN light in it) from a series of science images containing the SN flux we wish to measure. For every science epoch, we align each of the underlying \webb Level 2 (CAL) images to a catalog of JADES galaxies (that has in turn been aligned to {\em Gaia} sources), using the \textit{JWST}/\textit{HST} Alignment Tool \citep[{\tt JHAT};][]{rest_arminrestjhat_2023}\footnote{\url{https://jhat.readthedocs.io}}. CAL images are those that have been bias-subtracted, dark-subtracted, and flat-fielded but not yet corrected for geometric distortion. We drizzle these CAL files into Level 3 (I2D) files using the \webb pipeline \citep{bushouse_jwst_2022}. The extra JHAT step improves the relative alignment by an order of magnitude between the epochs (from $\sim1$~pixel to $\sim0.1$~pixel), allowing for subtractions with fewer artifacts between the template and science images. To perform the subtraction, we use the High Order Transform of PSF and Template Subtraction \citep[{\tt HOTPANTS};][]{becker_hotpants_2015}\footnote{\url{https://github.com/acbecker/hotpants}} code \citep[with additional improvements implemented in {\tt photpipe};][]{rest_testing_2005}, resulting in the difference images upon which we perform our photometry (see Figure \ref{fig:im_cutouts}; the right three columns are the difference images [per filter] generated from subtracting the ``template'' [top, leftmost epoch] from the ``science'' image [bottom, leftmost image]).

We measure the photometry in the difference images with a process described in \citet{Pierel2024b,Pierel2024d}, using the \texttt{space\_phot}\footnote{\url{space-phot.readthedocs.io}} Level 3 point-spread function (PSF) fitting routine centered on a $5\times5$ pixel cutout at \highzII's position. \texttt{space\_phot} models the Level 3 PSF by drizzling the Level 2 PSF models from {\tt webbpsf}\footnote{\url{https://webbpsf.readthedocs.io}}, which account for the spatial and temporal dependence of the \webb PSF and corrects for the losses in flux incurred by imposing a finite aperture. The resulting fluxes, measured in units of MJy/sr, are converted to AB magnitudes using the native pixel scale of each image ($0.03\arcsec/$pix for short- and $0.06\arcsec/$pix for long-wavelength), and the final, measured photometry is given in Table \ref{tab:phot}.

\begin{table}
    \centering
    \caption{\label{tab:phot} Observations for \highzII discussed in Section \ref{sec:obs}.}
    
    \begin{tabular*}{\linewidth}{@{\extracolsep{\stretch{1}}}*{5}{c}}
\toprule
PID&MJD&Instrument&\multicolumn{1}{c}{Filter/Grating}&\multicolumn{1}{c}{m$_{AB}$}\\
1180&60216&NIRCam&F115W&30.04$\pm$0.12\\
1180&60216&NIRCam&F150W&28.83$\pm$0.06\\
1180&60217&NIRCam&F200W&28.07$\pm$0.04\\
1180&60216&NIRCam&F277W&27.94$\pm$0.04\\
1180&60217&NIRCam&F356W&27.99$\pm$0.05\\
1180&60216&NIRCam&F444W&28.14$\pm$0.06\\
\hline
1180&60276&NIRCam&F115W&>29.8\\
1180&60276&NIRCam&F150W&>29.5\\
1180&60276&NIRCam&F200W&28.49$\pm$0.13\\
1180&60276&NIRCam&F277W&28.17$\pm$0.12\\
1180&60276&NIRCam&F356W&28.03$\pm$0.11\\
1180&60276&NIRCam&F444W&28.25$\pm$0.21\\
\hline
6541&60310&NIRCam&F150W&>30.1\\
6541&60310&NIRCam&F200W&29.05$\pm$0.15\\
6541&60310&NIRCam&F277W&28.51$\pm$0.16\\
6541&60310&NIRCam&F356W&28.13$\pm$0.14\\
6541&60310&NIRCam&F444W&28.64$\pm$0.30\\
\hline
$6541$&$60310$&NIRSpec&Prism&--\\
\hline
\hline
    \end{tabular*}
\begin{flushleft}
\tablecomments{Columns are: \textit{JWST} Program ID, Modified Julian date, \webb instrument, NIRCam filter, and photometry plus final uncertainty for \highzII.  Upper limits are 5$\sigma$.}

\end{flushleft}
\end{table}

\subsection{NIRSpec Reduction}
\label{sub:obs_spec}
\begin{table}[h!]
    \centering
    \caption{\highzII NIRSpec Observation Details\label{tab:spec_details}}
    \begin{tabular}{ll}
        \hline
        Instrument & NIRSpec  \\
        Mode & MOS  \\
        Wavelength Range & 0.6 $-$ 5.3$\mu$m  \\
        Slit & 3 Shutter ($0.46'' \times 0.2''$ each)  \\
        Grating/Filter & Prism/CLEAR  \\
        R $=\lambda/\Delta\lambda$ & $\sim 30-300$ \\
        Readout Pattern & NRSIRS2  \\
        Groups per Integration & 19  \\
        Integrations per Exposure & 2  \\
        Exposures/Nods & 3  \\
        Total Exposure Time & 22,175s
    \end{tabular}
\end{table}

\begin{figure*}
    \centering
    \includegraphics[width=\textwidth,trim={0cm 0cm 0cm 0cm},clip]{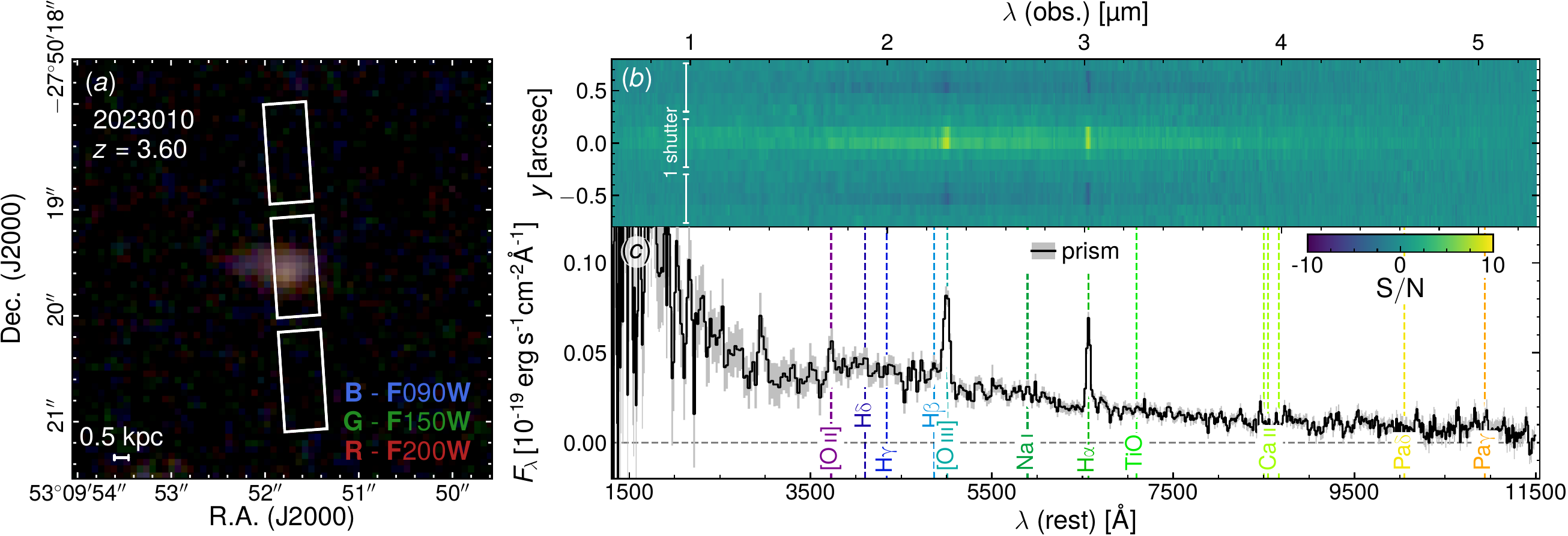}
    \caption{\textit{a:} The MSA slitlet position over \highzII. \textit{b:} The 2D NIRSpec Prism spectrum of \highzII and \highzIIhost.
    \textit{c:} The 1D-extracted NIRSpec spectrum for \highzII transformed into the rest-frame, with host emission lines color-coded and labeled. A spectroscopic redshift of $z=$\snz was measured based on the host's [O\,III] and H$\alpha$ lines. No SN features are readily apparent in the resulting 1D spectrum.}
    \label{fig:spec_2d_cuts}
\end{figure*}

We obtained the Stage 2 spectroscopic data collected from the DDT program from the Mikulski Archive for Space Telescopes (MAST; see Table~\ref{tab:spec_details}). With the context file {\tt jwst\_1185.pmap}, we used the \webb pipeline \citep{bushouse_jwst_2022} to generate the two-dimensional (2D) spectrum and applied a correction for slit-losses based on the position of the SN within the MSA shutters (Figure \ref{fig:spec_2d_cuts}, a and b). Next, we performed an optimal point source extraction using the algorithm from \citet[][implemented as a Jupyter notebook as part of the NIRSpec IFU Optimal Point Source Extraction guide\footnote{\url{https://spacetelescope.github.io/jdat_notebooks/notebooks/ifu_optimal/ifu_optimal.html}}]{horne_optimal_1986} to extract the superimposed spectra of the SN and its host. For an SN II, we expect H$\alpha$ to be the brightest feature during the photospheric phase, and in our last epoch (when the spectrum was obtained, see Table~\ref{tab:phot}) \highzII's flux in F277W (near 3~$\mu$m) is $\sim14$~nJy while the flux in the spectrum is $\sim50$~nJy at the same wavelength. With nearly $3/4$ of the flux contaminated with host light, even if a decomposition were possible, the SN spectrum is likely to have a signal-to-noise of $\sim3$ per pixel according to the \webb exposure time calculator\footnote{\url{https://jwst-docs.stsci.edu/jwst-exposure-time-calculator-overview}}. For these reasons, we use the Prism spectrum primarily to measure \highzII's redshift
(Figure \ref{fig:spec_2d_cuts}, c), and the oxygen line ratios to estimate the host's metallicity (see Section \ref{sub:host_props}). 





\subsection{Host Galaxy Redshift}
\label{sub:host_z}

\highzII was discovered in the host galaxy \highzIIhost, and the first step in analyzing \highzII is to determine its host redshift by identifying prominent emission lines seen in Figure \ref{fig:spec_2d_cuts}. These lines are best-matched by [O\,III] and H$\alpha$, which have rest-frame wavelengths of $\sim5008.24\textup{~\AA}$ and $\sim6564.61\textup{~\AA}$ in vacuum, respectively, and provide a robust spectroscopic redshift of $z_{\rm spec}=$\snz. We use this value for all analyses going forward, and present a detailed analysis of these host properties in Section~\ref{sub:host_props}.

\section{Analysis}
\label{sec:analysis}

\subsection{SN Light Curve Matching}
\label{sub:classification}

We fit the measured photometry from Table \ref{tab:phot} with the SALT3-NIR SN\,Ia light curve model \citep[][]{pierel_salt3nir_2022} and all existing CC\,SN light curve evolution models with rest-frame UV to near-IR (to observer-frame $\sim4~\mu$m) wavelength coverage \citep[][and references therein]{pierel_extending_2018}. These models are empirical spectral templates created from extremely well-observed, low-$z$ CC\,SNe and represent a wide range of diversity in each sub-type. In general, these spectral energy distributions (SEDs) are used to fit against our measured photometry, however, none of the templates extend to the rest-frame wavelengths covered by the F115W filter ($\sim2500\textup{~\AA})$, so it is excluded in the fitting (but see Section \ref{sub:lc_modeling}). We include Galactic dust based on the maps of \citet{schlafly_measuring_2011} and the reddening law from \citet{fitzpatrick_correcting_1999}, which corresponds to $E(B-V)=0.01$~mag with $R_V=3.1$. We also allow for host galaxy dust (up to $E(B-V)=1.5$~mag with $1<R_V<5$) of rest-frame, host-galaxy dust in the CC\,SN light curve fits and a SALT3-NIR color parameter range of $-1<c<1$. 

\begin{figure*}
    \centering
    \includegraphics[width=\textwidth]{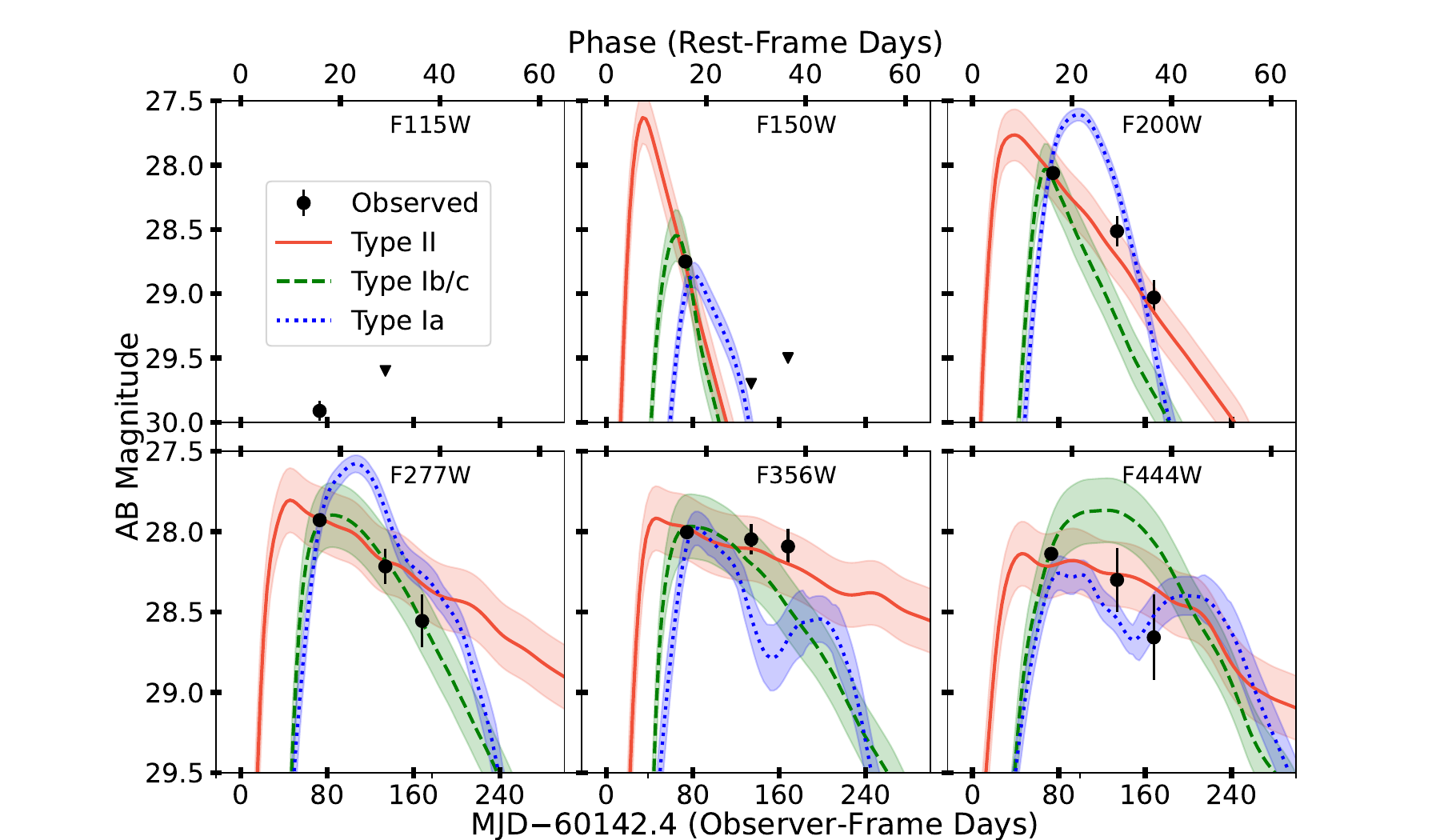}
    \caption{The photometry measured in Section \ref{sub:obs_phot} is shown as black circles with errors, with ($2\sigma$) upper-limits denoted by triangles. The best fit SN\,II (red solid line), SN\,Ib/c (green dashed line), and SN\,Ia (blue dotted line) models are shown for comparison. The SN\,II model shown is the SN\,2006kv template discussed in Section \ref{sub:classification}. The uncertainties shown are purely statistical.}
    \label{fig:lc_fit_cc}
\end{figure*}

Figure \ref{fig:lc_fit_cc} shows the best fit models for each SN sub-type in all filters. The resulting $\chi^2$ per degree of freedom ($\nu$, or reduced-$\chi^2$) for each model is shown in Table \ref{tab:chisq}. The SN\,Ia and SN\,Ib/c sub-types are heavily disfavored (best fit $\chi^2/\nu=16.63$ and $6.76$, respectively) compared to SN\,II ($\chi^2/\nu=1.10$). We take the results of this light curve fitting process as conclusive and give \highzII a classification of Type II as a result. The best fit SED to our photometry is that of SN\,2006kv, a normal SN\,IIP discovered at $z=0.0620$ \citep{dandrea_type_2010}. We note, however, that the UV coverage of SN\,2006kv's spectral template did not extend to cover \highzII's F115W detection (at $z=3.61$, F115W $\sim2500$~\AA; see Figure~\ref{fig:lc_fit_cc}), and therefore is omitted from the fitting. This blue emission could plausibly be due to a more exotic explosion with similarities to a SN\,II, a possibility that we explore in Section \ref{sub:lc_modeling}. While the fit to SN\,2006kv is quite good (see Table~\ref{tab:chisq}), \highzII's luminosity required a modeled peak $B$-band absolute magnitude of $-18.3\pm0.1$~mag, $\sim0.5$~mag brighter relative to the real SN\,2006kv; while this is still within the range of normal SN\,IIP absolute magnitudes observed in the local Universe ($\sim3\sigma$ above the distribution mean \citep{richardson_absolute_2014}), it is also in agreement with the suggestion from \citet{Scott2019} that low metallicity SN~II could be up to $\sim0.5$~mag brighter than SN~II at high metallicity.

\begin{table}
    \centering
    \caption{\label{tab:chisq} Comparison of the best-fit model $\chi^2$ statistic for each SN sub-type. }
    
    \begin{tabular*}{\linewidth}{@{\extracolsep{\stretch{1}}}*{3}{c}}
\toprule
SN Type&Mode/Template&$\chi^2/\nu$\\
\hline
Ia&SALT3-NIR&$16.63$\\
Ib/c&SDSS$004012$&$6.76$\\
II&SN\,2006kv&$1.10$\\
\end{tabular*}
\begin{flushleft}
\tablecomments{Columns are: SN type, spectral model/template used, and the light curve fitting $\chi^2$ per degree of freedom (DOF; $\nu$) \textit{without} model uncertainties, as they do not exist for CC\,SN models.}

\end{flushleft}
\end{table}



\subsection{Host Galaxy Properties}
\label{sub:host_props}



\begin{figure*}
    \centering
    \includegraphics[trim={0cm 35cm 0cm 0cm},clip,width=\textwidth]{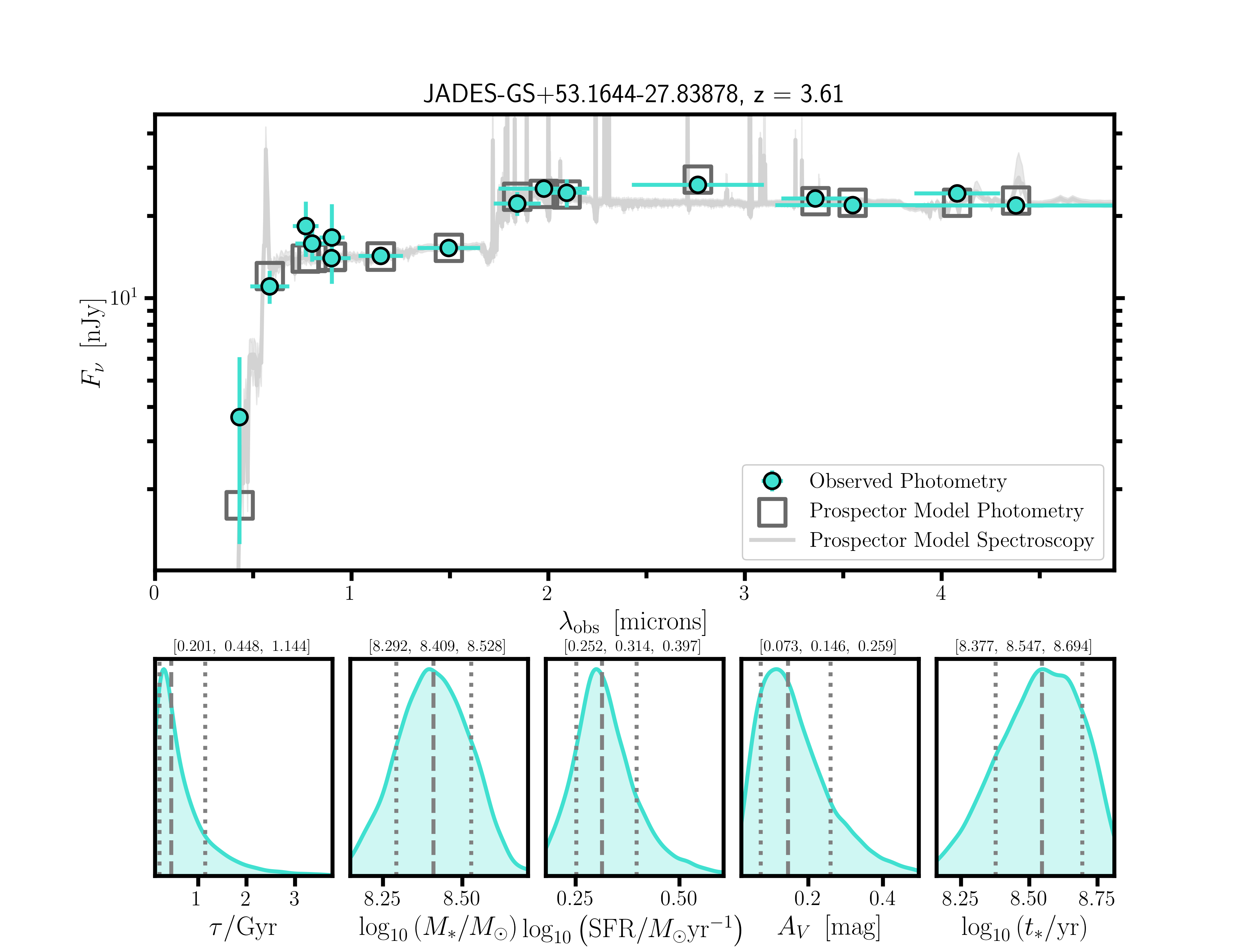}
    \caption{\label{fig:prospector_fit} Host template photometry fit using \texttt{Prospector}. The blue circles represent the observed JADES pre-SN photometry, and the dark grey shaded line represents the 50th percentile of the final {\tt Prospector} fit to the photometry, with a lighter grey color showing the 16th and 84th percentiles on the fit. The grey boxes are the estimated {\tt Prospector} photometry corresponding to the fit. We provide the  derived {\tt Prospector} host galaxy parameters in Table \ref{tab:tr10_host_results}.}
\end{figure*}

\def\arraystretch{1.25}
\begin{deluxetable}{l c c c}
\tablecaption{{\tt Prospector} Derived Host Properties
\label{tab:tr10_host_results}}
\tablehead{
    \colhead{Parameter} & \colhead{Value} 
}
\startdata
log(Age [$\mathrm{t_{*}/yr}$]) & \hostage \\
log(Stellar mass formed [M$_{*}$/M$_\odot$]) & \hostmass \\
log(SFR/[M$_\odot$yr$^{-1}$]) & \hostSFR \\
Gas-Phase Metallicity [Z$_\odot$] \tablenotemark{*} & \hostgasmetallicitysolar \\ 
log(O/H) + 12 \tablenotemark{$\rm{\dag}$} & \hostgasmetallicitylogOH \\
A$_\mathrm{v}$ [mag] &  \hostextinction \\
\enddata
\tablenotetext{*}{We derive a gas-phase metallicity using the oxygen line ratio diagnostic $\rm{O_{3}O_{2}}$ from \citet{Curti2020}; see Section~\ref{sub:host_props} for a detailed discussion.}
\tablenotetext{\dag}{We convert between gas-phase metallicity expressed in solar units to units of log(O/H) + 12 following the relation in \citet{Asplund2009}.}
\end{deluxetable}
\def\arraystretch{1.0}

At a redshift of $z=3.61$, the host of \highzII opens a window into the environment of a SN when the Universe was $<$ 2 Gyr old.
However, because there are no clear SN features in the spectrum for \highzII, yet we know that SN light must be contaminating the spectrum, any fit of the star formation history (SFH) of \highzIIhost will be biased by this unaccounted for SN light -- with the added light leading to systematically higher masses, and the SN color altering the inferred stellar properties. To address this, we perform a fit to the pre-SN photometry for the host galaxy to explore the SFH. We fit the JADES photometry for the source measured from the Hubble Space Telescope (HST) Advanced Camera for Surveys (ACS) in filters F435W, F606W, F775W, F814W, and F850LP along with JWST/NIRCam in the filters F090W, F115W, F150W, F182M, F200W, F210M, F277W, F335M, F356W, F410M, and F444W. For the fit, we use the tool {\tt Prospector}\footnote{\url{https://prospect.readthedocs.io/en/stable/}} \citep{prospector2021} and follow the method outlined in Helton et al. (in preparation). Briefly, within {\tt Prospector} we employ the Flexible Stellar Population Synthesis (FSPS) code \citep{conroy2009, conroy2010}, and we sampled the posterior distributions of the stellar population properties using the dynamic nested sampling code \texttt{dynesty}\footnote{\url{https://dynesty.readthedocs.io/en/stable/}} \citep{speagle2020}. We utilize a Chabrier initial mass function (IMF) with a lower bound of 0.08 $M_{\odot}$ and an upper bound of 120 $M_{\odot}$. Additionally, we assume a delayed-$\tau$ star-forming history of the form $\mathrm{SFR} \sim t_{\mathrm{age}} \times e^{-t_{\mathrm{age}} / \tau}$, where SFR is the star formation rate, $t_{\mathrm{age}}$ is the age of the galaxy, and $\tau$ is the e-folding time. For the fit, we fix the redshift to $z = 3.61$, and allow the stellar- and gas-phase metallicity to vary uniformly between log(Z/Z$_\odot$) $= -3.0 - 0.0$. We plot the {\tt Prospector} fit corresponding to the 50th percentile on the posterior, along with the fit photometry, in Figure~\ref{fig:prospector_fit}. 

From these fits we estimate a host mass of $\mathrm{log_{10}}(M_{*}/M_{\odot})=$~\hostmass, host age $\mathrm{log_{10}}(t_{*}/yr)=$~\hostage, and host extinction $A_{v}=$~\hostextinction~mag.
These and additional host properties are summarized in Table~\ref{tab:tr10_host_results}. 

Because there are no clear SN features in the spectrum for \highzII, we rely on the metallicity inferred from the host to estimate the metallicity of the SN. However, while we use {\tt Prospector} to infer host properties like mass, we do not use it to infer the host metallicity because the host SED modeling can be unreliable due to the strong degeneracy between metallicity and stellar age \citep{Dotter2017}. To infer the host metallicity, we instead turn to spectral fitting of the forbidden oxygen lines present in the spectrum (see Figure~\ref{fig:spec_2d_cuts}). In the photospheric phase, we do not expect much SN contamination in [\ion{O}{2}] and [\ion{O}{3}], and use the ratio of [\ion{O}{3}] to [\ion{O}{2}] (i.e., the $\rm{O_{3}O_{2}}$ diagnostic from \citet{Curti2020}) to estimate the metallicity at the position of the SN. We find that $\rm{O_{3}O_{2}}=3.0^{+3.2}_{-1.1}$, and assuming a solar metallicity of $\mathrm{log(O/H)}+12=8.69$ \citep{Asplund2009}, we find a host oxygen abundance of $\mathrm{log(O/H)}+12=8.1\pm0.2$, or Z$_{*} \approx 0.3$~Z$_\odot$. We note that {\tt Prospector} finds a gas-phase metallicity of $\mathrm{log(O/H)}+12=7.1\pm0.1$, or Z$_{*} \approx 0.02$~Z$_\odot$, $\sim$ an order of magnitude lower. This is a substantial discrepancy, however, we adopt the derivation from the oxygen ratio due to the aforementioned issues when using the integrated fit from {\tt Prospector}. For the remainder of the paper, we adopt a gas-phase metallicity of $\mathrm{log(O/H)}+12=8.1\pm0.2$. This metallicity is notably lower than the mean derived oxygen metallicity found for a collection of SNe II (dominated by IIP) by \citet{Anderson2010} of $\mathrm{log(O/H)}+12=8.580\pm0.027$.




\begin{figure*}[th!]
    \centering
    \includegraphics[width=0.8\textwidth]{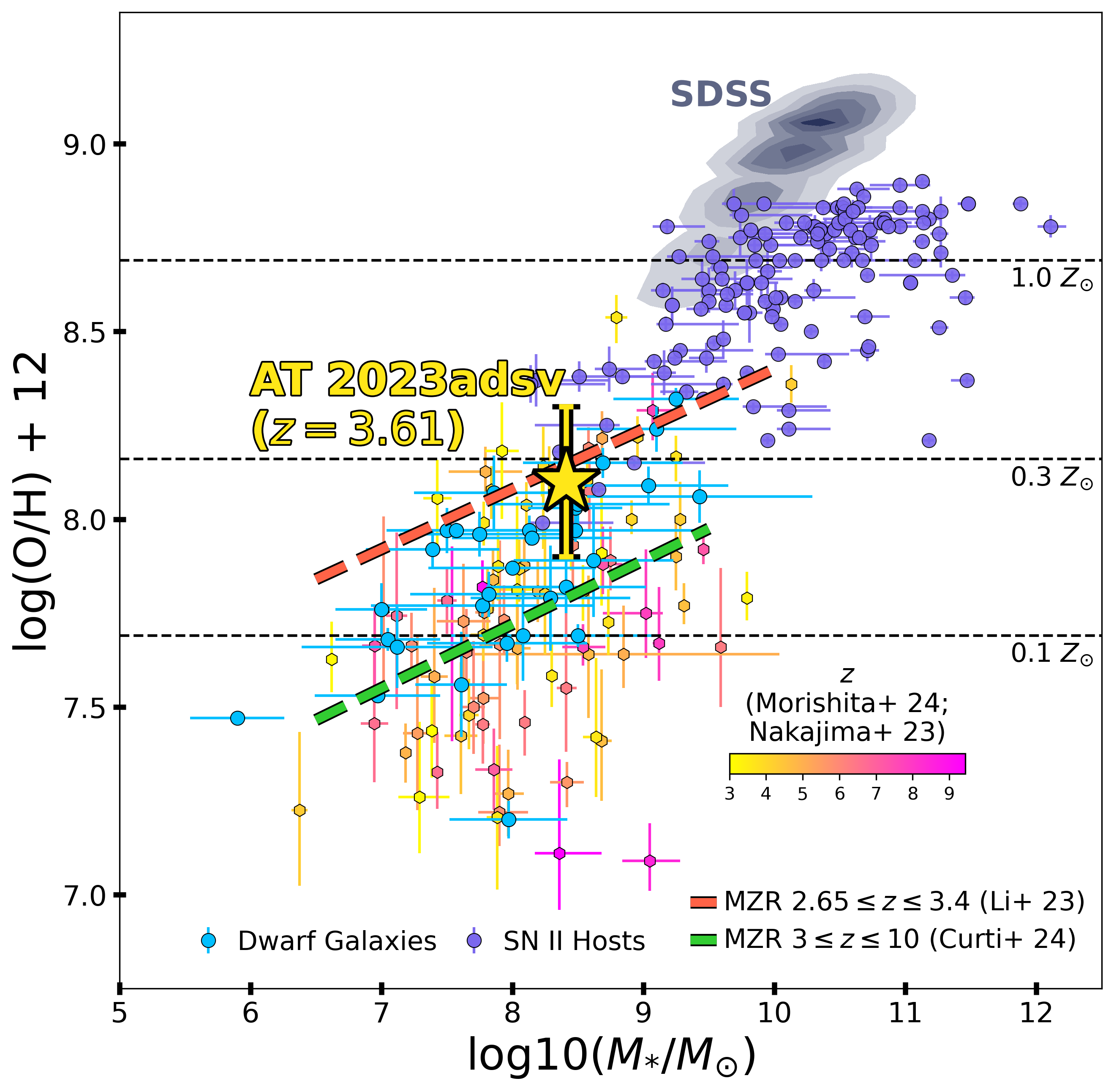}
    \caption{\highzII's inferred host galaxy mass and metallicity (gold star) compared to a selection of local galaxies. Grey contours correspond to galaxies selected from SDSS DR8 with $z < 0.7$ \citep{Aihara2011, Eisenstein2011}, purple points correspond to core-collapse SN hosts \citep{kelly_core_2012}, yellow-pink points correspond to \webb -selected galaxies redshifts at $3 \leq z \leq 9$ \citep{Nakajima2023, Curti2024, Morishita2024}, and blue points correspond to low-metallicity dwarf galaxies from \citet{Berg2012}. The red dashed line corresponds to the mass-metallicity relationship (MZR) for galaxies at $2.65 \leq z \leq 3.4$ from \citet{Li2023}; the green dashed line corresponds to the MZR at $3 \leq z \leq 10$ from \citet{Curti2024}. Overplotted are horizontal dashed lines which correspond to the the $1.0$, $0.3$, and $0.1$ solar oxygen abundance values derived from \citet{Asplund2009}.}
    \label{fig:host_comparisons}
\end{figure*}

We place this galaxy in a wider context of SNe~II hosts in Figure \ref{fig:host_comparisons}. We compare the mass and metallicity of the host of \highzII with a population of $z < 0.7$ galaxies from SDSS DR8 \citep[grey contours; ][]{Aihara2011, Eisenstein2011}, galaxies from \webb with redshifts $3 \leq z \leq 9$ \citep[yellow-pink points;][]{Nakajima2023, Morishita2024, Curti2024}, core-collapse SN hosts \citep[purple points;][]{kelly_core_2012}, and low-metallicity dwarf galaxies \citep[blue points;][]{Berg2012}. Metallicities for both the SN hosts and SDSS sample were derived following the PP04 O3N2 calibration from \citet{Pettini2004}, while metallicities for both the \webb-selected, high-$z$ sample and dwarf sample were derived using the direct electron-temperature method \citep{Campbell1986}. Over-plotted are horizontal dashed lines corresponding to the the $1.0$, $0.3$, and $0.1$ solar oxygen abundance values converted from \citet{Asplund2009}, as well as the mass-metallicity relationship (MZR) for galaxies at $2.65 \geq z \geq 3.4$ from \citet[][red dashed line]{Li2023}, and the MZR  at $3 \geq z \geq 10$ from \citet[][green dashed line]{Curti2024}. These MZR scalings are supported by recent work with \webb \citep{Schaerer2022, Taylor2022, Katz2023, Rhoads2023}, tracing this relation to even further distances with measurements of two galaxies at $z \approx 8$, and has confirmed that at fixed stellar mass, galaxies are generally less enriched at higher redshift \citep{Langeroodi2023}. We find that the metallicity of the host of \highzII is consistent with the MZR from \citet{Li2023} as well as with the lower-metallicity tail of the core-collapse distribution.


\subsection{Light curve Modeling}
\label{sub:lc_modeling}
In order to estimate the explosion properties of \highzII, we compare synthetic light curves with those of \highzII. For this purpose, we first obtained red supergiant (RSG) SN progenitor models with 0.3~Z$_{\odot}$ (in agreement with the inferred metallicity measured for its host; see Section~\ref{sub:host_props}) by using \texttt{Modules for Experiments in Stellar Astrophysics (MESA)} version r23.05.1 \citep{paxton2011mesa,paxton2013mesa,paxton2015mesa,paxton2018mesa,paxton2019mesa,jermyn2023mesa}. 
We selected a grid of models with zero-age main-sequence (ZAMS) masses ($\mathrm{M_{ZAMS}}$) of 12, 16, and 20~M$_{\odot}$. The details of the assumptions in the stellar evolution calculations are presented by the accompanying paper \citep{Moriya2025}.
The final progenitor properties are summarized in Table~\ref{tab:progenitor}.

The RSG progenitor models are then transferred to the one-dimensional multi-frequency radiation hydrodynamics code \texttt{STELLA} \citep{blinnikov1998,blinnikov2000,blinnikov2006}. \texttt{STELLA} numerically evaluates the SED evolution of SNe, and thus we can directly estimate light curves in the observer frame from the theoretical SED evolution when they appear at $z=3.6$. We refer to the accompanying paper for the details on the light curve calculations \citep{Moriya2025}.
Because SNe~II are generally found to be embedded within a dense and confined circumstellar medium \citep[CSM; e.g.,][]{forster2018}, we also include a version of each of our models with this close in CSM (deposited up to $10^{15}~\mathrm{cm}$). This approach was taken to account for the amount of UV flux detected in the first epoch (i.e., at $z=3.6$, F115W and F150W span $\sim2500-3250$~\AA~in the rest-frame), and the fact that a typical mass-loss rate of $10^{-3}~\mathrm{M_\odot~yr^{-1}}$ with a wind velocity of $10~\mathrm{km~s^{-1}}$, can act as an additional early power source in the light curve \citep{Moriya2011, Dessart2022}. The confined CSM mass is $0.07~\mathrm{M_\odot}$ ($M_\mathrm{ZAMS}=12\mathrm{M_\odot}$), 
$0.11~\mathrm{M_\odot}$ ($M_\mathrm{ZAMS}=16\mathrm{M_\odot}$), and $0.16~\mathrm{M_\odot}$ ($M_\mathrm{ZAMS}=20\mathrm{M_\odot}$). 

\begin{figure*}[th!]
    \centering
    \includegraphics[width=0.99\textwidth]{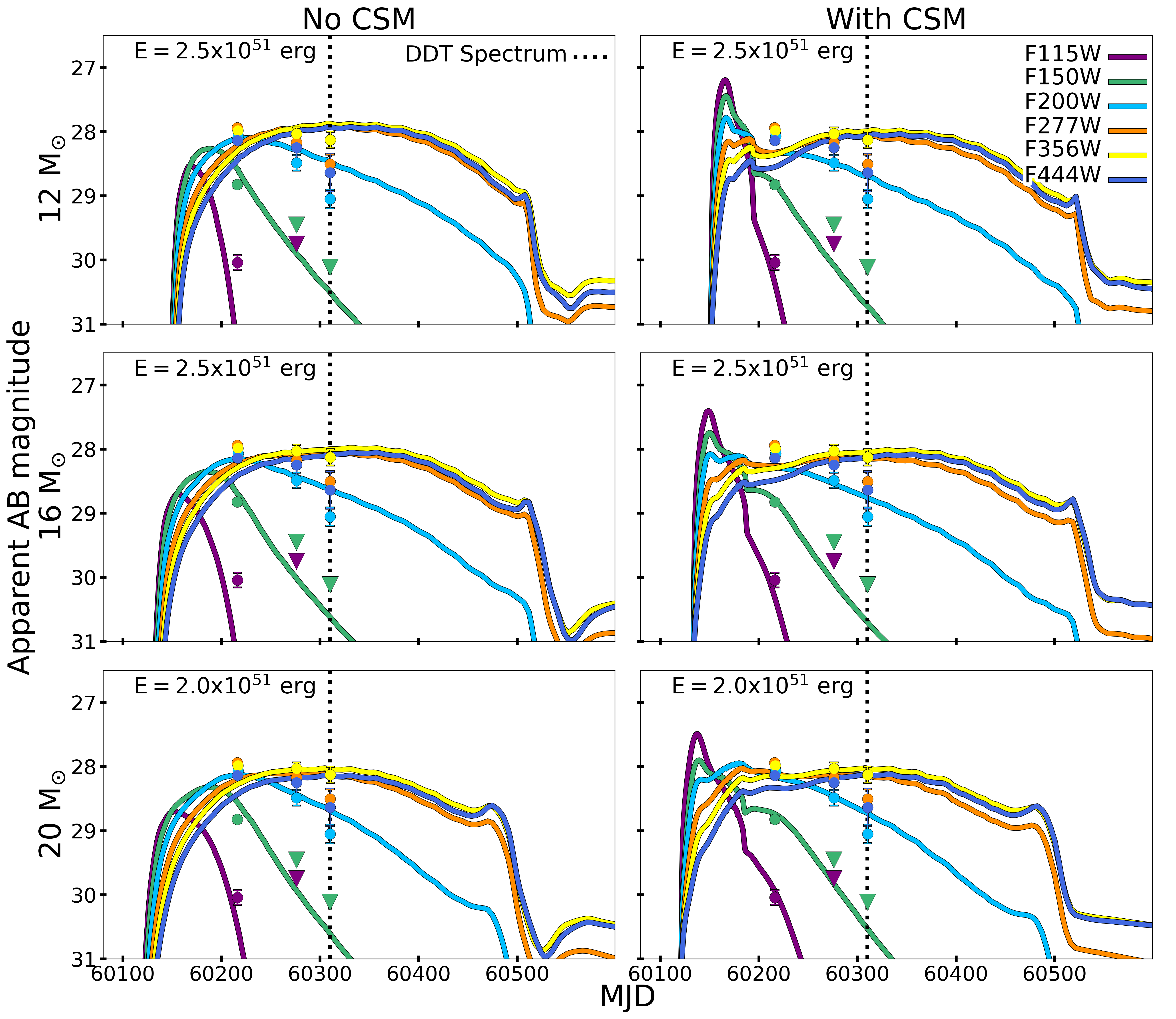}
    \caption{A grid of light-curve models for \highzII based on the 0.3~Z$_\odot$ RSG SN progenitors used in this study. Each row corresponds to a different ZAMS mass (from the top to bottom row: 12, 16, and 20~M$_\odot$) with the left column free from confined CSM, and the right column containing a confined CSM mass of $0.07~\mathrm{M_\odot}$ ($12~\mathrm{M_\odot}$), $0.11~\mathrm{M_\odot}$ ($16~\mathrm{M_\odot}$), and $0.16~\mathrm{M_\odot}$ ($20~\mathrm{M_\odot}$) at a radius of $10^{15}~\mathrm{cm}$. We vary the explosion energy between $2.0-3.0\times 10^{51}~\mathrm{erg}$. The higher mass models better match the range of observations of \highzII than the models with $M<20M_\odot$, although we cannot distinguish between the models with and without CSM. A more complete discussion is presented in Section \ref{sub:lc_modeling}. The vertical dashed line represents the time of the DDT spectrum. }
    \label{fig:rsg_models}
\end{figure*}

Figure~\ref{fig:rsg_models} presents the results of the numerical modeling. We find that an explosion energy of $(2-3)\times 10^{51}~\mathrm{erg}$ is required to account for the brightness of \highzII. While the three progenitor models explain the overall properties of \highzII well, the $12\mathrm{M_\odot}$ and $16\mathrm{M_\odot}$ models without CSM struggle to reproduce the observed UV flux in the first epoch. In the models with CSM, both the $12\mathrm{M_\odot}$ and $16\mathrm{M_\odot}$ models do better at matching the early-time UV flux, but are underluminous compared to the first detections in F277W, F356W, and F444W despite boosting their explosion energies to $2.5\times 10^{51}$~ergs. All models fail to fit the last epoch F200W, F277W, and F444W detections. Despite this, we find that the best overall fit to be the $20\mathrm{M_\odot}$ progenitors (both with and without CSM), and cannot distinguish between them due to not having observations during the timeframe of the inferred modeled peak.

\subsubsection{Pair-Instability Explosion}
\label{sub:lc_modeling_pisn}

\begin{figure}[th!]
    \centering
    \includegraphics[width=0.49\textwidth]{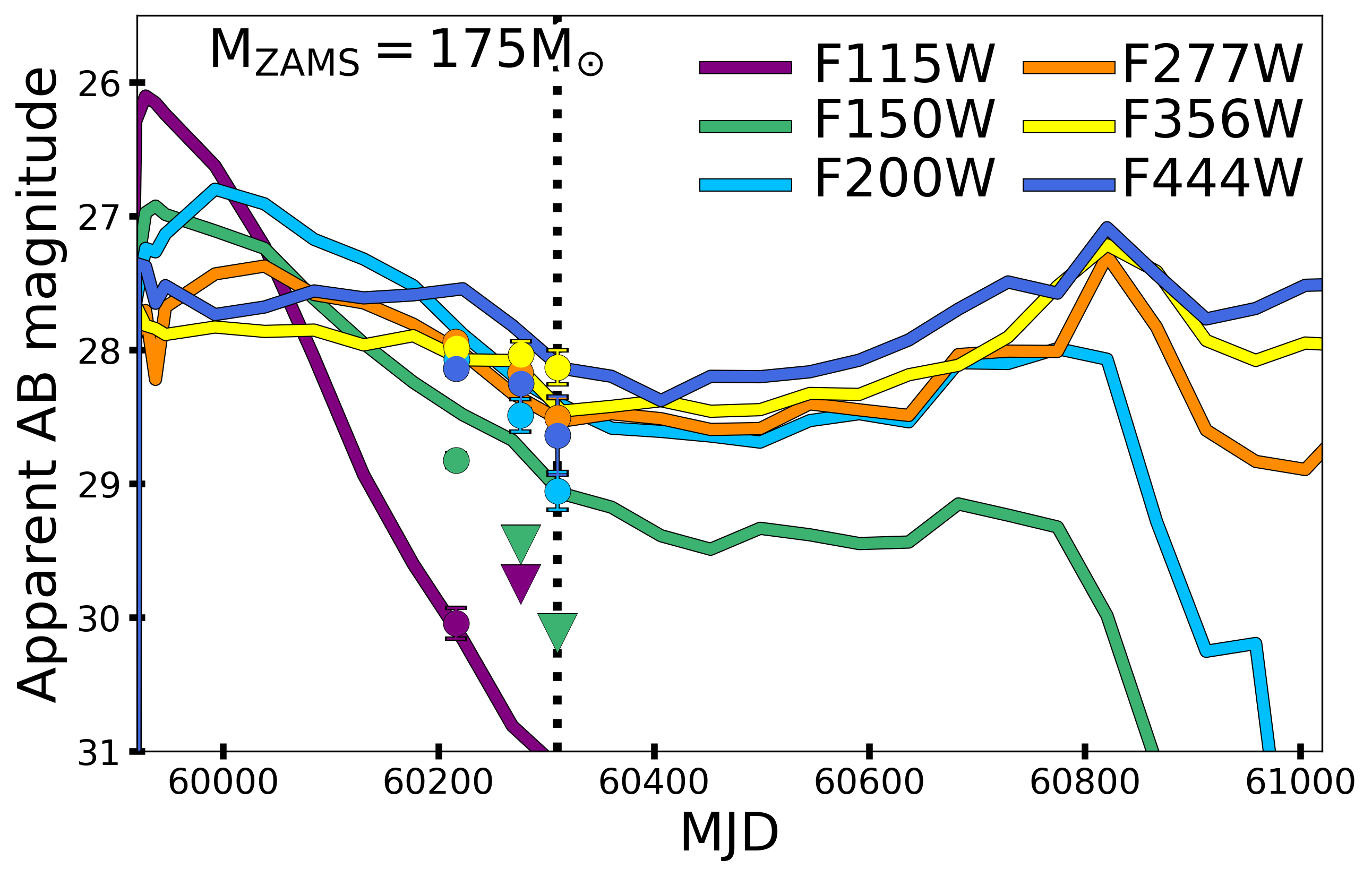}
    \caption{\highzII compared with the RSG PISN model (R175) from \citet{kasen_pair_2011}. While the PISN model fit is reasonable for \highzII, long term monitoring of this object would be needed to distinguish whether \highzII is a typical SNe or a more exotic PISN.}
    \label{fig:pisn_model}
\end{figure}

Motivated by exploring an alternative explanation for the measured early blue flux at the redshift of \highzII, as well as the prediction that PISNe could form at metallicities as high as $\mathrm{Z_\odot}/3$ \citep[the metallicity of \highzII's host;][]{Langer2007}, we compared PISN light-curve models from \citet{kasen_pair_2011} to our measured photometry. We found that the 175~M$_{\odot}$ RSG PISN model (R175) matches well to \highzII, but that the dataset as it stands is not sufficient to differentiate between a PISN model and those explored in (Fig.~\ref{fig:pisn_model}). The PISN has a much longer duration than typical RSG explosions discussed before, so more observations at later times would be required to confirm (or rule out) a PISN progenitor. Because of the expected low PISN event rates, \highzII is likely to be a RSG explosion. However, future wide-area surveys will be able to discover PISNe and high-$z$ SNe~II such as \highzII, demonstrating the importance of long-term monitoring of the same field in order to distinguish between PISNe and other typical SNe.

\subsection{Color-Magnitude Comparison}
\label{sub:color_mag}

\begin{figure*}
    \centering
    \includegraphics[width=\textwidth]{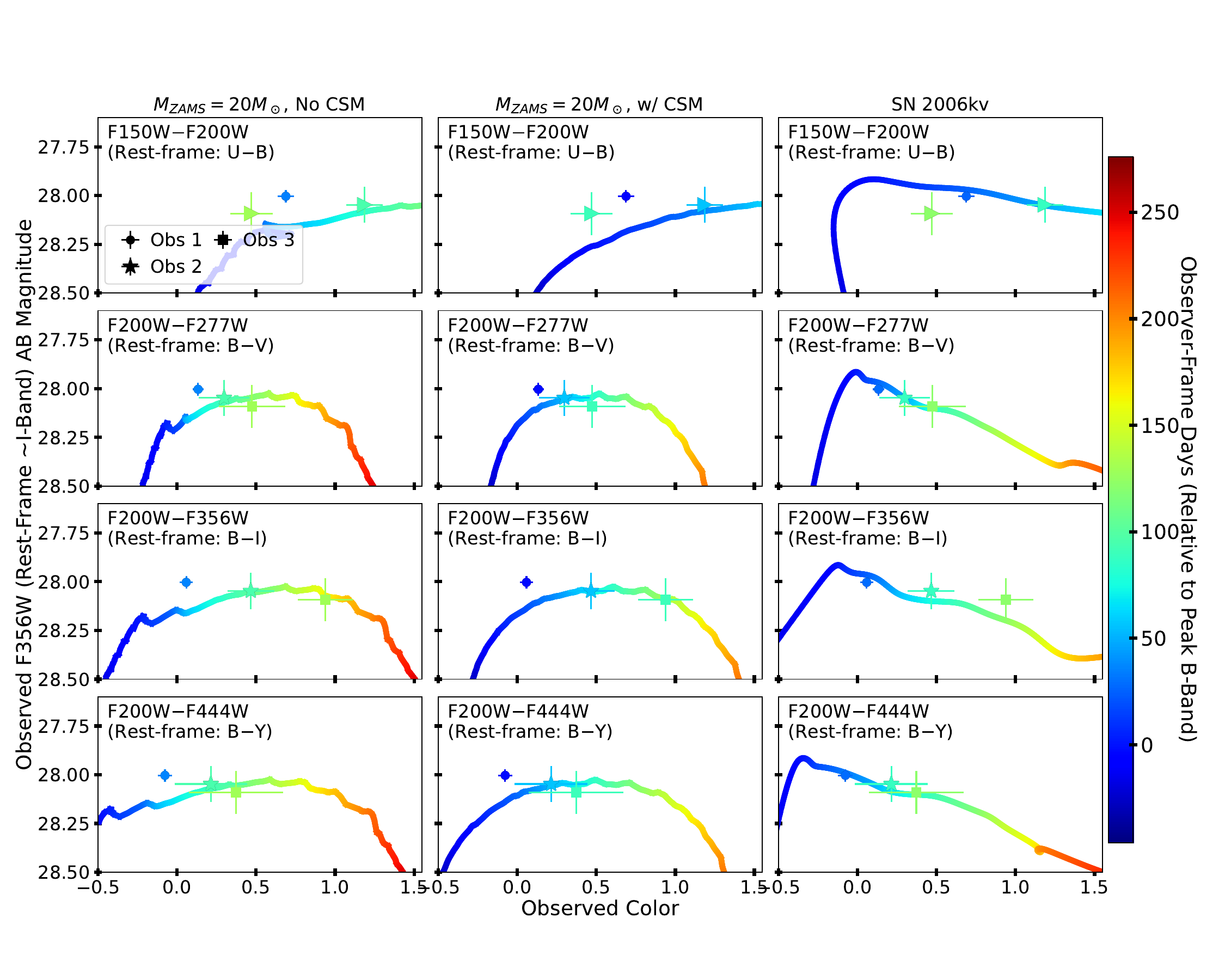}
    \caption{Four observed colors (labeled by row) vs. magnitude (F356W, rest-frame $\sim$I-band) shown in the legend as black points with error bars, with the symbols corresponding to the three observed epochs (a legend in upper-left; order of observations is circle, star, square), and limits shown with directional arrows. The colored lines (and the fill color of the symbols) track the corresponding color-magnitude space as a function of time from best fit models, with MESA models from Section \ref{sub:lc_modeling} in the first two columns and the best spectral template fit (SN\,2006kv) in the third column (see Section \ref{sub:classification}). The coloring of the lines is described by the color bar (right), with early times shown as blue and late times as red.  }
    \label{fig:col_mag_phase}
\end{figure*}

In Figure \ref{fig:col_mag_phase} we plot the observed colors vs observed F356W (rest-frame $\sim I$-band) magnitude of the best RSG models ($20~\mathrm{M_\odot}$ with and without CSM interaction), along with SN\,2006kv, a normal SN\,IIP and the best fit spectral template match found in Section~\ref{sub:classification}. The shape of each marker corresponds to the epoch of the observation, with the circle, square, and star corresponding to the first, second, and third observed epochs listed in Table~\ref{tab:phot}. Triangles are upper-limit measurements. The colored lines track the corresponding color-magnitude evolution of each model as a function of time, where the color of the line reflects the observer-frame days relative to the $B$-band peak for these models. From this evolution plot, SN\,2006kv shows a striking similarity. We leverage this similarity when comparing \highzII to a collection of local SNe~IIP in Section~\ref{sub:lowz_comp}.


\begin{table}
    \centering
    \caption{\label{tab:progenitor} Progenitor properties for light-curve computations.}
    
    \begin{tabular*}{\linewidth}{@{\extracolsep{\stretch{1}}}*{4}{c}}
\toprule
$M_\mathrm{ZAMS}$ & $M_\mathrm{fin}$& $M_\mathrm{H-rich}$ & $R_\mathrm{fin}$ \\
($\mathrm{M_\odot}$) & ($\mathrm{M_\odot}$)& ($\mathrm{M_\odot}$)& ($\mathrm{R_\odot}$) \\
\hline
12& 11.8 &  8.6 & 434 \\
16& 14.7 & 10.0 & 632 \\
20& 15.9 &  9.5 & 847 \\
175 & 163.8 & 79.4  & 2499  \\
\end{tabular*}
\begin{flushleft}
\tablecomments{Columns are: ZAMS mass ($M_\mathrm{ZAMS}$), final mass at explosion ($M_\mathrm{fin}$), hydrogen-rich envelope mass at explosion ($M_\mathrm{H-rich}$), and progenitor radius at explosion ($R_\mathrm{fin}$).}

\end{flushleft}
\end{table}




\section{Discussion}
\label{sec:discussion}

\subsection{\highzII's Metallicity and Comparison to Low-$z$ SN IIPs}
\label{sub:lowz_comp}

\begin{figure}
    \centering
    \includegraphics[width=0.49\textwidth]{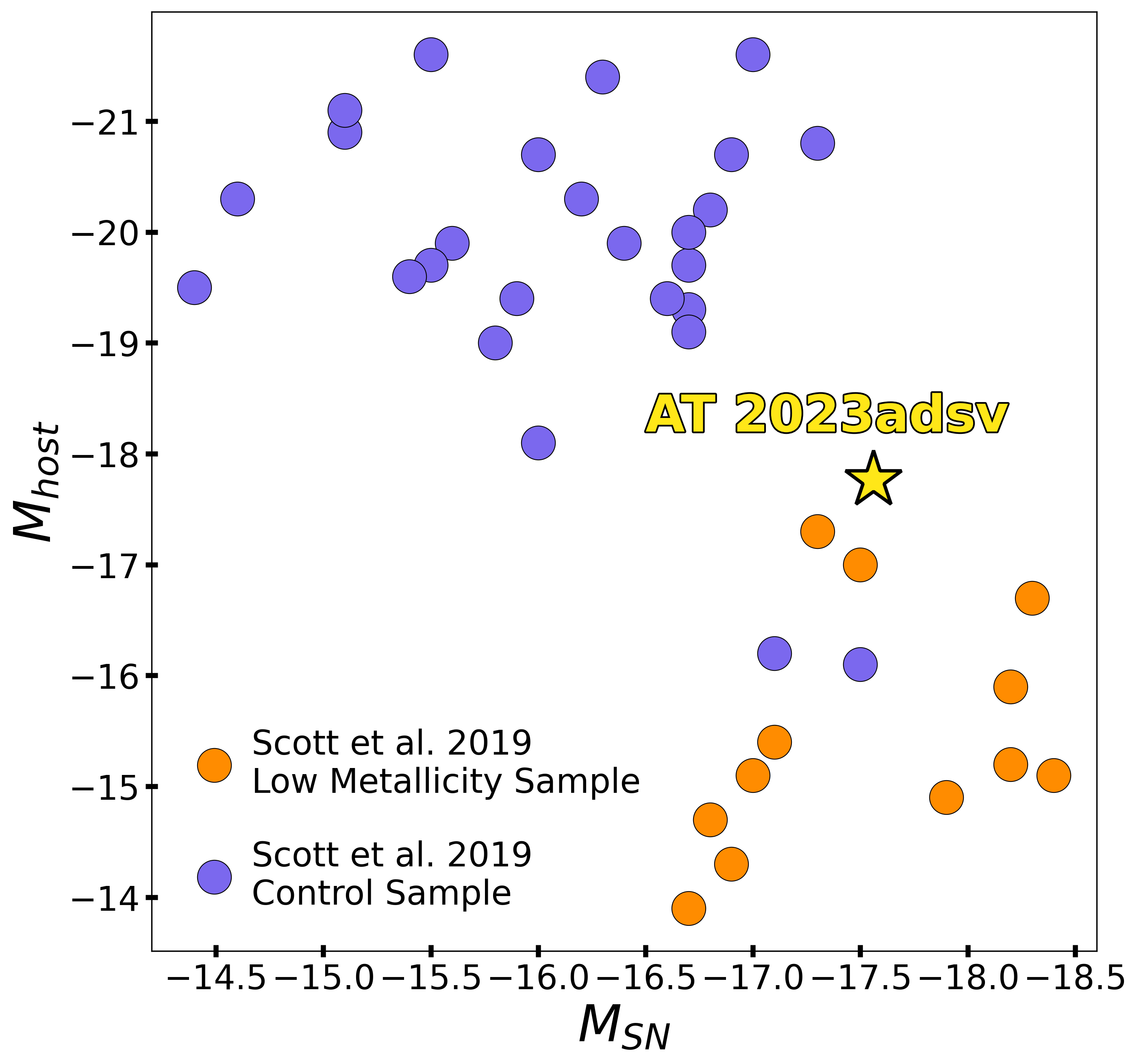}
    \caption{Figure adapted from \citet{Scott2019}, with a sample of local, low-metallicity selected SN~IIP (orange points) compared to a ``control'' sample of local, SN~IIP from the literature (purple points). All photometry is reported in $r$-band, and SN luminosities are taken during the plateau phase. The gold star corresponds to an estimate of \highzII's plateau luminosity were it an SN~IIP (see Section~\ref{sub:lowz_comp}) along with its host luminosity, which occupies a space between the two samples. }
    \label{fig:snII_vs_hosts}
\end{figure}

The question of metallicity is a central one when considering the likely explosion scenario for \highzII, as well as in understanding the physical origin of its bright UV luminosity in the first \webb epoch. In particular, higher metallicities can lead to more pre-SN mass loss via line-driven stellar winds \citep{Mokiem2007}, a lower mass hydrogen envelope (and therefore a shorter duration plateau phase of a SN IIP), and a smaller ejecta mass of the explosion. Depending on the location, density, and distribution of this wind-driven material, the SN shock breakout from core-collapse can be extended from a baseline of $1$-$1000$ seconds to many hours \citep{Gezari2015, forster2018}, and at high-$z$ these same effects could span {\it days} in the observer-frame. In general, this interaction between the shock and the surrounding CSM can lead to an increased UV and optical luminosity in the observed light curve \citep{Schlegel1990, Moriya2011}. Lower metallicity, on the other hand, will lower the opacity of the envelope, resulting in hotter and more compact stars. This lower opacity decreases the line blanketing of the UV portion of the spectrum, allowing more blue light to escape the SN \citep{Eastman1994, Dessart2013}. 

However, indirectly ascertaining an SN IIP's metallicity from its light curve, e.g. by studying its luminosity and color evolution, is further complicated by the degeneracy of this evolution with the effect that the progenitor's radius has on its explosion properties. Specifically, during the SN explosion, the expansion from a smaller radius contributes to faster cooling and therefore faster color evolution during the photospheric phase; this radius is in turn sensitive to effects like stellar rotation, convective overshoot, and mixing (see \citet{Dessart2013} for a review). For these reasons, and owing to the fact that the outermost ejecta of SNe IIP have photospheres which are characterized by the molecular clouds from which the pre-SN stars are formed \citep{Dessart2014}, definitive assessments of an SN IIP's metallicity are done through detailed spectroscopic studies of the strengths of metal-line absorptions during an SN~IIP's plateau phase \citep{Dessart2020}. In particular, the pEW of Fe II $\lambda$5018~\AA~has been shown as a proxy for the metallicity of an SN~IIP, with larger widths suggesting higher metallicities. This result has been confirmed observationally in the local universe \citep{Anderson2016, Taddia2016, Gutierrez2017, Gutierrez2018}, but relies on spectral coverage during the plateau phase to accurately measure the effect. 

In our analysis of \highzII, we do not have a spectrum with distinct SN features to perform this measurement, but an intriguing study by \citet[][hereafter, S19]{Scott2019} finds that when comparing the plateau luminosity of a sample of SN~IIP with their host luminosity (in rest-frame $r$-band), spectroscopically-confirmed, low metallicity SNe~IIP tend to separate from a ``control'' sample taken from the literature. Specifically, S19's sample are constructed from ``high-contrast'' SN~IIP --- those with high SN luminosities but with low luminosity hosts, versus a sample of SN~IIP without this property. The central idea is to test the assertion that these high-contrast SN~IIP also have low metallicities as measured in spectra taken during their plateau phase (i.e., because of the inferred low metallicity of low luminosity hosts via the MZR). S19's result confirms this statistically, and crucially, S19 compares the low-metallicity sample with the control sample photometrically. 

Motivated by this, and because \highzII is well-modeled by SN\,2006kv, we fit the last observed epoch of \highzII (F200W, F277W, F356W, and F444W) with a blackbody and find a temperature of $\sim6200$~K --- within the range of the recombination temperatures of $H$ which power the plateau-phase of an SN~IIP's light curve \citep[$5500-7000$~K;][]{Dessart2014, Dessart2020}. Therefore, if we assume that \highzII~{\it is} an SN~IIP and is within or {\it near} the plateau-phase in the last observed \webb epoch, we can provisionally compare \highzII with the sample in S19. We perform $k$-corrections on both the host and SN photometry, and find that the fitted SN\,2006kv model's luminosity at $+50$~days is $-17.5$~Mag in $r$-band. The resulting comparison is shown in Figure~\ref{fig:snII_vs_hosts}.

In this parameter space, \highzII's high luminosity places its abscissa in the same region as the low-metallicity SNe~IIP selected by S19, however, \highzII's host brightness places its ordinate between the control sample and the low-metallicity sample. The apparent tension of an ``overluminous'' host for a lower-metallicity SN~IIP can be related to the the result presented in Figure~\ref{fig:host_comparisons}. In the context of an MZR that evolves with redshift (see Section~\ref{sub:host_props}), we expect that at a fixed stellar-mass, O/H decreases with increasing redshift, which for a host at $3 \leq z \leq 4$ would move \highzII's host luminosity closer to the control (i.e., higher metallicity) sample.

\section{Conclusion}
\label{sec:conclusion}


We have presented \webb observations of \highzII with a spectroscopic redshift of $z=$\snz, which we classify using 
light curve information as a relatively bright ($M_B=-18.3\pm0.1$mag) SN\,II. We further model the light curve using the 
\texttt{MESA} code and find a good match to a RSG progenitor star with ZAMS mass $20M_\odot$, albeit with a slightly high explosion energy of $2.0 \times 10^{51}$~ergs. \highzII could also plausibly be a PISN, but such a rare object is a less likely explanation than a normal SN\,II, and more observations with a longer temporal baseline would be necessary to confirm or rule out such a result. 


While the DDT spectrum of \highzII confirms its redshift, we could not reliably isolate specific SN features in the spectrum, and we limited our our analysis of \highzII and \highzIIhost to their photometric properties to prevent bias. Examining the host of \highzII, \highzIIhost, we find a relatively low-mass ($\mathrm{log_{10}}(M_{*}/M_{\odot})=$~\hostmass), moderately dusty ($A_{v}=$~\hostextinction~mag), low-metallicity (Z$_{*}$~=~\hostgasmetallicitysolar~Z$_\odot$) galaxy. A more careful study of the host environment, and potentially of the SN itself, would be possible with a template spectrum of the host with the same observing parameters in a future \webb cycle -- allowing a direct comparison of a spectrum with, and without, contaminating SN light.

\highzII is likely the most distant SN\,II with a spectroscopic redshift yet found \citep[although see AT~2023adst, a reported SN~II with a host $z=4.117$, yet with a much less robust classification;][]{decoursey_jades_2024},
and provides a timely opportunity to study massive SN progenitors at $z>3$. Intriguingly, \highzII's inferred metallicity places it in a parameter space between that of low-$z$, low-metallicity SNe~IIP and a control sample of solar metallicity SNe~IIP hosted in massive galaxies; it will be necessary to continue to observe such distant CC\,SNe with \webb to statistically test whether such CC\,SNe are indeed well-modeled by massive and metal poor progenitors with higher than average explosion energies. Carefully following up these high-$z$ SNe may lead to novel constraints on the early Universe IMF, as well as metal enrichment and mixing. Upcoming surveys such as the \textit{Nancy Grace Roman Space Telescope} High Latitude Time Domain Survey \citep[HLTDS;][]{hounsell_simulations_2018,rose_reference_2021} will likely open a new frontier for this science by finding thousands of distant massive progenitor CC\,SNe. However, \webb will remain our only resource capable of rest-frame optical-IR imaging and spectroscopy at high-$z$, highlighting the need for building a sample of such observations now and into the future.






\begin{center}
    \textbf{Acknowledgments}
\end{center}

This paper is based in part on observations with the NASA/ESA/CSA {\it Hubble Space Telescope} and {\it James Webb Space Telescope} obtained from the Mikulski Archive for Space Telescopes at Space Telescope Science Institute (STScI), 
which is operated by the Association of Universities for Research in Astronomy, Inc. (AURA), under NASA contract NAS 5-03127 for {\it JWST}. These observations are 
associated with program \#1180 and \#6541. The {\it JWST} data used in this paper can be found in MAST: \dataset[10.17909/h522-xx30]{https://dx.doi.org/10.17909/h522-xx30}.
We thank the DDT and {\it JWST}/{\it HST} scheduling teams at STScI for their extraordinary effort in getting the DDT observations used here scheduled quickly. 
The specific observations analyzed can be accessed via \dataset[DOI: 10.17909/snj9-an10]{https://doi.org/10.17909/snj9-an10}; support was provided to JDRP and ME through program HST-GO-16264. JDRP is supported by NASA through a Einstein
Fellowship grant No. HF2-51541.001 awarded by STScI, which is operated by AURA,
for NASA, under contract NAS5-26555.
Numerical computations were in part carried out on PC cluster at the Center for Computational Astrophysics, National Astronomical Observatory of Japan. TJM is supported by the Grants-in-Aid for Scientific Research of the Japan Society for the Promotion of Science (JP24K00682, JP24H01824, JP21H04997, JP24H00002, JP24H00027, JP24K00668) and by the Australian Research Council (ARC) through the ARC's Discovery Projects funding scheme (project DP240101786). AJB acknowledges funding from the ``FirstGalaxies'' Advanced Grant from the European Research Council (ERC) under the European Union's Horizon 2020 research and innovation program (Grant agreement No. 789056). DJE is supported as a Simons Investigator and by {\it JWST}/NIRCam contract to the University of Arizona, NAS5-02015. BDJ acknowledges the {\it JWST}/NIRCam contract to the University of Arizona, NAS5-02015. RM acknowledges support by the Science and Technology Facilities Council (STFC), by the ERC through Advanced Grant 695671 ``QUENCH'', and by the UKRI Frontier Research grant RISEandFALL. RM also acknowledges funding from a research professorship from the Royal Society. BER acknowledges support from the NIRCam Science Team contract to the University of Arizona, NAS5-02015, and {\it JWST} Program 3215. ST acknowledges support by the Royal Society Research Grant G125142. QW is supported by the Sagol Weizmann-MIT Bridge Program. The authors acknowledge use of the lux supercomputer at UC Santa Cruz, funded by NSF MRI grant AST 1828315.

\clearpage

\bibliographystyle{aasjournal}


\end{document}